\documentclass[conf]{new-aiaa}
\usepackage[utf8]{inputenc}

\usepackage{graphicx}
\usepackage{amsmath}
\usepackage[version=4]{mhchem}
\usepackage{siunitx}
\usepackage{longtable,tabularx}
\usepackage{adjustbox}

\usepackage{longtable,tabularx}
\usepackage{amsmath} % assumes amsmath package installed
\usepackage{amsfonts}
\usepackage{amsthm}

\theoremstyle{definition}

\usepackage{float} 
\usepackage[skip=0em,font=footnotesize]{caption}
\usepackage[margin = 1in]{geometry}

\usepackage{subcaption}

\usepackage{tikz}
\usetikzlibrary{shapes,arrows,calc,positioning}
\tikzstyle{bigblock} = [draw, fill=blue!20, rectangle, 
    minimum height=6em, minimum width=8em]
\tikzstyle{medblock} = [draw, fill=blue!20, rectangle, 
    minimum height=4em, minimum width=4em]    
\tikzstyle{mux} = [draw, fill=black!20, rectangle, 
    minimum height=5em, minimum width=0.1em]    
\tikzstyle{smallblock} = [draw, fill=blue!20, rectangle, 
    minimum height=2em, minimum width=3em]
    
\tikzstyle{data_block} = [draw, fill=green!20, rectangle, 
    minimum height=2em, minimum width=3em]
\tikzstyle{ops_block} = [draw, fill=blue!20, rectangle, 
    minimum height=2em, minimum width=3em]    
\tikzstyle{est_block} = [draw, fill=red!20, rectangle, 
    minimum height=2em, minimum width=3em]    
    
\tikzstyle{sum} = [draw, fill=blue!20, circle, node distance=1cm,minimum height=0.5cm]
\tikzstyle{signal} = [coordinate]
\tikzstyle{pinstyle} = [pin edge={to-,thin,black}]
\tikzstyle{block} = [draw, fill=blue!20, rectangle, 
    minimum height=3em, minimum width=9em]
\tikzstyle{blockS} = [draw, fill=blue!20, rectangle, 
    minimum height=3em, minimum width=4em]    
\tikzstyle{input} = [coordinate]
\tikzstyle{output} = [coordinate]
\usetikzlibrary{matrix}

\usetikzlibrary{positioning}
\usetikzlibrary{math}

\usepackage{hyperref}
\usepackage{xcolor}
\hypersetup{
    colorlinks,
    linkcolor={blue!100!black},
    citecolor={blue!50!black},
    urlcolor={blue!80!black}
}

\newcommand{\bc}{\begin{center}}
\newcommand{\ec}{\end{center}}
\newcommand{\benum}{\begin{enumerate}}
\newcommand{\eenum}{\end{enumerate}}
\newcommand{\nn}{\nonumber}
\newcommand{\matl}{\left[ \begin{array}}
\newcommand{\matr}{\end{array} \right]}

\renewcommand{\matl}{\begin{bmatrix}}
\renewcommand{\matr}{\end{bmatrix}}

\newcommand{\matls}{\left[ \begin{smallmatrix}}
\newcommand{\matrs}{\end{smallmatrix} \right]}
\newcommand{\isdef}{\stackrel{\triangle}{=}}

\newcommand{\inv}{^{-1}}

\newcommand{\vect}[1]{\overset{\rightharpoonup}{#1}}

\newcommand{\rmA}{{\rm A}}

\newcommand{\rmC}{{\rm C}}
\newcommand{\rmD}{{\rm D}}

\newcommand{\rmL}{{\rm L}}
\newcommand{\rmM}{{\rm M}}

\newcommand{\rmS}{{\rm S}}
\newcommand{\rmT}{{\rm T}}

\newcommand{\rma}{{\rm a}}

\newcommand{\rmg}{{\rm g}}

\newcommand{\rmi}{{\rm i}}

\newcommand{\rmr}{{\rm r}}
\newcommand{\rms}{{\rm s}}

\newcommand{\SN}{{\mathcal N}}

\usepackage{enumitem,amssymb}
\newlist{todolist}{itemize}{2}
\setlist[todolist]{label=$\square$}
\usepackage{pifont}

\newcommand{\ihat}[1]{\ensuremath{\hat \imath_{\rm #1} }}
\newcommand{\jhat}[1]{\ensuremath{\hat \jmath_{\rm #1} }}
\newcommand{\khat}[1]{\ensuremath{\hat k_{\rm #1} }}
\usepackage[colorinlistoftodos,textwidth=1 in,textsize=footnotesize]{todonotes}
\usepackage{tikz}
\usetikzlibrary{shapes,arrows,calc,positioning}
\usetikzlibrary{calc,patterns,decorations.pathmorphing,decorations.markings,positioning}
\tikzset{dashed/.style={dash pattern=on 4pt off 1.5pt,line width=0.5pt},
	wing/.style={smooth,line width=1pt,fill=black!5},
	point/.style={circle,inner sep=0pt, minimum width=4pt,fill=black}
}

\tikzstyle{spring}=[thick,decorate,decoration={zigzag,pre length=0.3cm,post length=0.3cm,segment length=6}]

\newcommand\spiral{} % just for safety
\def\spiral[#1](#2)(#3:#4)(#5:#6)[#7]{%
	\pgfmathsetmacro{\domain}{#4+#7*360}
	\pgfmathsetmacro{\growth}{180*(#6-#5)/(pi*(\domain-#3))}
	\draw [#1,
	shift={(#2)},
	domain=#3*pi/180:\domain*pi/180,
	variable=\t,
	smooth,
	samples=int(\domain/5)] plot ({\t r}: {#5+\growth*\t-\growth*#3*pi/180})
}

\newcommand\intrado{(0,0)
	(-0.0334,-0.0767)
	(-0.1087,-0.1437)
	(-0.2253,-0.2011)
	(-0.3824,-0.2489)
	(-0.5790,-0.2870)
	(-0.8139,-0.3158)
	(-1.0860,-0.3355)
	(-1.3940,-0.3466)
	(-1.7365,-0.3497)
	(-2.1123,-0.3457)
	(-2.5199,-0.3356)
	(-2.9580,-0.3209)
	(-3.4252,-0.3029)
	(-3.9198,-0.2835)
	(-4.4427,-0.2625)
	(-4.9936,-0.2377)
	(-5.5666,-0.2102)
	(-6.1594,-0.1810)
	(-6.7696,-0.1513)
	(-7.3950,-0.1217)
	(-8.0332,-0.0930)
	(-8.6815,-0.0653)
	(-9.3376,-0.0386)
	(-9.9988,-0.0125)}
\newcommand\extrado{(0,0)
	(-0.0095,0.0831)
	(-0.0624,0.1691)
	(-0.1590,0.2574)
	(-0.2990,0.3467)
	(-0.4824,0.4357)
	(-0.7085,0.5225)
	(-0.9765,0.6050)
	(-1.2855,0.6812)
	(-1.6341,0.7488)
	(-2.0206,0.8055)
	(-2.4433,0.8492)
	(-2.8998,0.8778)
	(-3.3879,0.8897)
	(-3.9049,0.8833)
	(-4.4459,0.8592)
	(-5.0064,0.8210)
	(-5.5876,0.7687)
	(-6.1870,0.7023)
	(-6.8016,0.6219)
	(-7.4286,0.5277)
	(-8.0650,0.4197)
	(-8.7080,0.2980)
	(-9.3544,0.1623)
	(-10.0012,0.0125)}

\newcommand{\frameVectrix}[1]{
\matl  
    \ihat{#1} \\
    \jhat{#1} \\
    \khat{#1} \\
\matr
}
\newcommand{\Otwo}[1]
{\ensuremath{
    \matl
        \cos #1 & 0 & -\sin #1 \\
        0 & 1 & 0 \\
        \sin #1 & 0 & \cos #1 
    \matr
}}

\title{A Review of Modeling and Controlling Flutter}
\title{Modeling and Data-Driven Flutter Suppression in a \\ Pitch--Plunge Aeroelastic System}

\title{Data-Driven Flutter Suppression in a \\ Pitch--Plunge Aeroelastic System}

\author{
Maximilian Moody\footnote{Undergraduate Student, Department of Mechanical Engineering, {\tt\small mmoody3@umbc.edu}}
, 
Olvin Moran\footnote{Undergraduate Student, Department of Mechanical Engineering, {\tt\small omoran1@umbc.edu}}
,
Juan Augusto Paredes Salazar\footnote{Postdoctoral Research Fellow, Department of Mechanical Engineering, {\tt\small japarede@umbc.edu}}
, 
Ankit Goel\footnote{Assistant Professor, Department of Mechanical Engineering, AIAA Member, {\tt\small ankgoel@umbc.edu}}
}
\affil{University of Maryland, Baltimore County, Baltimore, MD 21250}

\begin{document}

\maketitle

% \begin{enumerate}
%     \item Section 3 - add Lassale statement
%     \item Section 4 - Expanbd section A and section B with commentary
%     \item Section 4 - add legends to Fig 2 Fig  3 7 and 8
%     \item Fig 10 - set ylim from 0 to 40
%     \item subscript d should be nonitalic
% \end{enumerate}

\begin{abstract}
This paper presents a first-principles nonlinear model of a pitch--plunge aeroelastic system and investigates data-driven control for flutter suppression.
The model is derived directly from Newtonian mechanics and coupled with a simplified aerodynamic force parameterization, enabling explicit representation of key physical and aerodynamic parameters.
A linearization-based stability analysis is used to characterize the onset of flutter and investigate the sensitivity of the stability boundary to variations in the lift-curve slope and freestream velocity.
Dynamic Mode Adaptive Control (DMAC) is employed to suppress aeroelastic vibrations using online local system identification and LQR feedback synthesis without requiring prior knowledge of the system dynamics.
Numerical simulations demonstrate flutter suppression and plunge reference tracking beyond the flutter boundary, where the performance of DMAC is compared with that of a model-based LQR.
Simulations with a varying freestream velocity further demonstrate that DMAC maintains reference tracking across the flutter boundary without being reset or retuned.
The results demonstrate the potential of DMAC for adaptive flutter stabilization in the pitch--plunge aeroelastic system.
\end{abstract}

% \keywords{one, two, three, four}
\textit{\bf Keywords:} aeroelasticity, flutter suppression, adaptive control, data-driven control.

% To be submitted to

% Track: Structural Dynamics  

% Subtopic: Flutter and Limit-Cycle Oscillations or Vibration and Vibroacoustic Control, Energy Harvesting, and Damping

%%%%%%%%%%%%%%%%%%%%%%%%%%%%%%%%%%%%%%%%%%%%%%%%%%%%%%%%%%%%%%%%%%%%%%%%%%%%%%%%
\section{Introduction}

Aeroelastic phenomena arise from the interaction between aerodynamic forces, structural elasticity, and inertial effects, and are a critical concern in a wide range of engineering systems including aircraft wings, wind turbine blades, and turbomachinery.
In such systems, flow-induced vibrations can lead to flutter, a dynamic instability characterized by sustained or growing oscillations that can severely limit performance, reduce structural lifespan, and, in extreme cases, lead to catastrophic failure.
The importance of understanding and mitigating aeroelastic effects has been well documented across applications in energy and aerospace systems \cite{wang2016state,fleeter1979aeroelasticity}.

Due to the strong coupling between fluid and structural dynamics, aeroelastic systems are inherently complex and sensitive to operating conditions and modeling assumptions.
Uncertainty in aerodynamic coefficients, structural parameters, and flow conditions further complicates analysis and control design \cite{pettit2004uncertainty}.
As a result, a significant body of work has focused on both modeling and active control strategies for flutter suppression, including model-based control designs and experimental demonstrations \cite{roger1975active,peloubet1984recent,mukhopadhyay2000transonic}, as well as robust and multivariable approaches for flexible aircraft and unmanned systems \cite{theis2016robust,theis2020robust,waszak2001robust,schmidt2016stability}.
Additional efforts have explored the role of control surfaces and aerodynamic nonlinearities in shaping aeroelastic response \cite{bueno2022control,kotikalpudi2018flutter,allen2003investigation}.
Adaptive control methods have also been investigated for flutter suppression, including adaptive feedback linearization for a typical wing section with structural nonlinearity \cite{ko1999adaptive}.
More recently, data-driven approaches have been considered, such as a model-free adaptive controller based on ridge regression of input and output data \cite{yu2024modelfree}, as well as a combination of dynamic mode decomposition with control and model predictive control for a pitch--plunge section \cite{mendez2026datadriven}.

Benchmark models such as the NASA Benchmark Active Controls Technology (BACT) configuration have played an important role in advancing the study of aeroelastic control \cite{waszak1998modeling}.
In this work, we develop a simplified nonlinear pitch--plunge aeroelastic model derived directly from Newtonian mechanics, coupled with a tractable aerodynamic force parameterization, where the parameter values are adapted from the BACT configuration.
This first-principles formulation enables explicit identification of key physical and aerodynamic parameters and provides a clear framework to study their influence on stability and dynamic response.

The resulting model serves as a testbed to investigate the performance of Dynamic Mode Adaptive Control (DMAC), a fully data-driven control framework that constructs local models online and synthesizes feedback directly from measured data \cite{oveissi2025model,oveissi2026dmac}.
The objective is to assess the ability of DMAC to suppress vibration and mitigate aeroelastic instability while following commanded plunge displacements, without requiring prior knowledge of the system dynamics.
The performance of DMAC is compared with that of an LQR designed using the linearized model, and is further examined under a freestream velocity that varies across the flutter speed.

The remainder of the paper is organized as follows.
Section~\ref{sec:modeling} presents the first-principles nonlinear model of the pitch--plunge aeroelastic system, including the aerodynamic force parameterization and state-space formulation.
Section~\ref{sec:stability_analysis} provides a linearization-based stability analysis to characterize the onset of flutter and assess parametric sensitivity.
Section~\ref{sec:control} investigates the application of Dynamic Mode Adaptive Control (DMAC) for flutter suppression.
Section~\ref{sec:numerical} presents numerical simulations to evaluate the performance of DMAC in suppressing flutter and following plunge commands.
Section~\ref{sec:conclusions} concludes the paper with a summary of the results.

\section{Modeling}
\label{sec:modeling}

This section derives the equations of motion of the pitch--plunge airfoil shown in Figure~\ref{fig:wing_diagram}.

\subsection{Geometry}
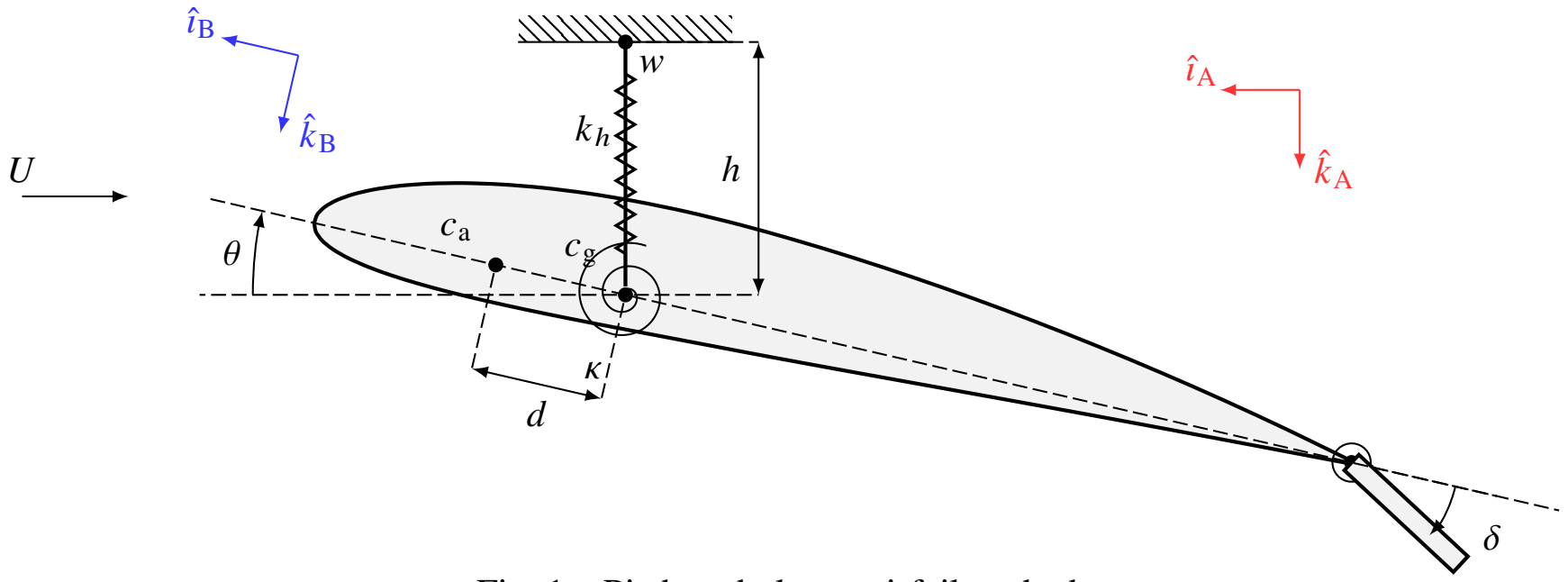
\begin{figure}[ht!]
    \centering
    \resizebox{0.7\columnwidth}{!}{%

            \begin{tikzpicture}[>=latex,scale=1,line width=0.5pt]
        \begin{scope}[xscale=-1, every node/.style={transform shape=false}]

        \begin{scope}[rotate around={13:(10,0)}]
            \draw[wing] plot coordinates {\intrado};
            \draw[wing] plot coordinates {\extrado};
            \coordinate(frameBody) at (0.5,1.5);
            \coordinate(frameBodyjhat) at (0.3,1.7);
            \draw[->,blue!80] (frameBody) -- +(0:0.75) node[above, near end, xshift = -1em, yshift = -0.25em] {$\hat{\imath}_{\rm B}$};
            \draw[->,blue!80] (frameBody) -- +(-90:0.75) node[right, near end, yshift = -0.5em] {$\hat{k}_{\rm B}$};
            % \draw[blue!80] (frameBodyjhat) circle (0.4em) node[xshift = -0.85 em, yshift = 0.85em] {$\hat{\jmath}_{\rm B}$};
            % \draw[blue!80, fill = blue!80] (frameBodyjhat) circle (0.15em);
            % points
            \node[point] at (-3,0) (EC) {};
            \node[point] at (-1.75,0) (AC) {};
            \draw[dashed] (EC) -- +(-90:1)node[coordinate](ec){};
            \draw[dashed] (AC) -- +(-90:1)node[coordinate](ec2){};
            \node[above left, outer sep=4pt] at (EC) {$c_{\rm g}$};
            \node[above left, outer sep=3pt] at (AC) {$c_{\rm a}$};
            \spiral[](EC)(0:100)(0:0.5)[2] node[coordinate] (SpringEnd) {};
            % \draw[fill=white,white] (-8.25,0) -- +(0:0.5) arc (0:360:0.5);
            % \draw[line width=1pt] (-8.25,0) -- +(60:.5) arc (60:-15:.5);
            % \fill[fill=white] (-8,-.25) rectangle (-10,0.5);

            % ---- Trailing-edge flap removed ----
            % (Deleted: winglet drawing, k_delta torsional spring, l_delta dimension, delta angle marker)
            % ---- Trailing-edge flap ----
            \coordinate (FH) at (-10,0);   % flap hinge near left end
            \def\lf{1.4}                     % flap length
            \def\tf{0.20}                    % flap thickness
            \def\deltaang{30}                   % flap angle from chord line [deg]
            
            % hinge
            \node[point] at (FH) {};
            \draw (FH) circle (0.18);
            
            % reference/chord line for flap angle
            \draw[dashed] (FH) -- ++(180:\lf);
            
            % flap body
            \begin{scope}[shift={(FH)}, rotate={180+\deltaang}]
                \draw[wing]
                    (0,\tf/2)
                    -- (\lf,\tf/2)
                    -- (\lf,-\tf/2)
                    -- (0,-\tf/2)
                    -- cycle;
            \end{scope}
            
            % delta angle
            \draw[->]
                (FH) +(180:1)
                arc[start angle=180,end angle={180+\deltaang},radius=1];
            
            \node at ($(FH)+(180+\deltaang/2:1.5)$) {$\delta$};
            % -------------------------------------

            \draw[dashed](1,0)--(-12,0);
        \end{scope}%
        \coordinate(frameInertia) at (-9,-1);
        \coordinate(frameInertiajhat) at (-9.2,-0.8);
        \draw[->,red!80] (frameInertia) -- +(0:0.75) node[above, near end, xshift = -1em, yshift = -0.25em] {$\hat{\imath}_{\rm A}$};
        \draw[->,red!80] (frameInertia) -- +(-90:0.75) node[right, near end, yshift = -0.5em] {$\hat{k}_{\rm A}$};
        % \draw[red!80] (frameInertiajhat) circle (0.4em) node[xshift = -0.85 em, yshift = 0.85em] {$\hat{\jmath}_{\rm A}$};
        % \draw[red!80, fill = red!80] (frameInertiajhat) circle (0.15em);
        %
        \draw[<->] (ec) -- node[below, yshift = -0.1em]{$d$}(ec2);
        \draw[->] (EC) +(0:3.5cm) arc [start angle=0, delta angle=13, radius=3.5cm] node[midway,left]{$\theta$};
        \draw[dashed] (EC) -- +(180:-4);
        \node[yshift=2.5cm,pattern=north west lines,minimum width=2cm] (Ground) at (EC) {};
        \draw[line width=1pt] (EC)-- node[right, xshift = -1.5em, yshift = -5.5em]{$\kappa$} (Ground.south);
        \draw (Ground.south west) -- (Ground.south east);
        \node[point] at (Ground.south) (gnd) {};
        \node[below right] at (gnd) {$w$};
        \draw [spring] (Ground.south) -- node[xshift = -0.85em, yshift = 0.1em,above] {$k_{h}$} (EC);
        \draw[dashed] (Ground.south) -- +(-180:1.25) node[coordinate](epsh){};
        \draw[dashed] (EC) -- +(-180:1.25) node[coordinate](epsh2){};
        \draw[<->] (epsh) -- node[left, xshift = -0.1em, yshift = 0.0em] {$h$}(epsh2);

        \draw[->] (3,-2) node [above] {$U$} -- +(-1,0)  ;
        \end{scope}
        
        \end{tikzpicture}

    }
    \caption{Pitch and plunge airfoil testbed.}
    \label{fig:wing_diagram}
\end{figure}

Let $\rm F_A$ be an inertial frame, and let $\rm F_B$ be a frame fixed to the wing as shown in Figure~\ref{fig:wing_diagram}.
The frame $\rm F_B$ is obtained by rotating $\rm F_A$ about $\jhat \rmA$ by the \textit{pitch angle} $\theta,$ which implies that the frame axes satisfy
\begin{align}
    \frameVectrix{B} &= \Otwo{\theta} \frameVectrix{A},
    \label{eq:frame_relations}
\end{align}
and
\begin{align}
    \vect \omega_{\rm B/A} = \dot \theta \jhat A.
\end{align}
A positive value of $\theta$ corresponds to pitch up.
The deflection $\delta$ of the trailing-edge flap is positive with the trailing edge down.
The linear and torsional spring deflections are
\begin{align}
    \tilde h &\isdef h-h_0,
    \quad
    \tilde \theta \isdef \theta -\theta_0,
\end{align}
where $h_0$ is the relaxed length of the linear spring and $\theta_0$ is the relaxed angle of the torsional spring.

\subsection{Kinematics}
Let $w$ be a point fixed in the ceiling to which the linear spring is connected.
Then,
\begin{align}
    \vect r_{c_\rmg/w}
        &=
            h \khat{A},
\end{align}
and thus the velocity and acceleration of $c_\rmg$ are
\begin{align}
    \vect v_{c_\rmg/w/\rmA}
        &=
            \dot h \khat{A},
    \\
    \vect a_{c_\rmg/w/\rmA}
        &=
            \ddot h \khat{A}
        =
            \ddot h (-\sin \theta \ihat{B} + \cos \theta \khat{B} ).
\end{align}
The aerodynamic center $c_\rma$ is located at a distance $d$ ahead of $c_\rmg,$ that is, $\vect r_{c_\rma/c_\rmg} = d \ihat{B}.$
The velocity of $c_\rma$ is thus
\begin{align}
    \vect v_{c_\rma/w/\rmA}
        &=
            \dot h \khat{A} - d \dot \theta \khat{B}.
    \label{eq:v_ca}
\end{align}

\subsection{Forces and Moments}
The forces acting on the airfoil are the gravity force $m \vect g = mg \khat{A},$ the spring force $\vect f_{\rms/c_\rmg} = - k_h\tilde h \khat{A},$ the reaction force $\vect f_{\rmr/c_\rmg} = f_\rmr \ihat{A},$ the lift $\vect L = - L \khat{B},$ and the drag $\vect D = - D \ihat{B}.$
The reaction force constrains $c_\rmg$ to move along a vertical line.
The gravity, spring, and reaction forces are applied at $c_\rmg,$ and the lift and drag are applied at $c_\rma.$
The lift and drag are defined along the body axes, which approximates the wind-axis directions for small angle of attack.
The total force on the wing is thus
\begin{align}
    \vect f_{\rm total}
        &=
            m \vect g + \vect D + \vect L + \vect f_{\rms/c_\rmg} + \vect f_{\rmr/c_\rmg}
        \nn \\
        &=
            m g \khat{A} - D \ihat{B}  -L \khat{B} - k_h\tilde h \khat{A} + f_\rmr \ihat{A}.
\end{align}
Using \eqref{eq:frame_relations}, it follows that
\begin{align}
    \vect f_{\rm total}
        &=
            (m g - k_h\tilde h ) (-\sin \theta \ihat{B} + \cos \theta \khat{B})
            - D \ihat{B}
            - L \khat{B}
            + f_\rmr (\cos \theta \ihat{B} + \sin \theta \khat{B})
        \nn \\
        &=
            ( -D - (m g - k_h\tilde h )\sin \theta + f_\rmr \cos \theta ) \ihat{B}
            + (- L + (m g - k_h\tilde h ) \cos \theta + f_\rmr \sin \theta ) \khat{B}.
\end{align}
The moments on the airfoil relative to $c_\rmg$ are due to the aerodynamic moment $\vect M_\rma = M_\rma \jhat A,$ the torsional spring with stiffness $\kappa,$ and the lift and drag.
The gravity, spring, and reaction forces are applied at $c_\rmg$ and thus do not contribute to this moment.
The total moment on the wing relative to $c_\rmg$ is thus
\begin{align}
    \vect M_{{\rm total}/c_\rmg}
        &=
            \vect M_\rma
            -
            \kappa \tilde \theta \jhat A
            +
            \vect r_{c_\rma/c_\rmg} \times (\vect D + \vect L)
        \nn \\
        &=
            M_\rma \jhat A
            -
            \kappa \tilde \theta \jhat A
            +
            d \ihat{B} \times (- D \ihat{B} - L \khat{B} )
        \nn \\
        &=
            (M_\rma - \kappa \tilde \theta + d L) \jhat A.
\end{align}

\subsection{Aerodynamic Force and Moment Parameterization}
\label{sec:aero_param}
The aerodynamic forces and moment are given by a quasi-steady model.
The model omits wake dynamics and aerodynamic memory effects.
Let $\rho$ denote the air density, $U$ the freestream speed, $S$ the wing planform area, and $c$ the chord length.
The lift, drag, and pitching moment about $c_\rma$ are parameterized in terms of the nondimensional coefficients $C_\rmL,$ $C_\rmD,$ and $C_\rmM$ as
\begin{align}
    L
        &=
            \tfrac12 \rho U^2 S  C_\rmL(\alpha, \delta),
    \label{eq:L_param}
    \\
    D
        &=
            \tfrac12 \rho U^2 S  C_\rmD(\alpha, \delta),
    \label{eq:D_param}
    \\
    M_\rma
        &=
            \tfrac12 \rho U^2 S c  C_\rmM(\alpha, \dot \theta, \delta),
    \label{eq:M_param}
\end{align}
where $\alpha$ is the effective angle of attack at $c_\rma.$
The freestream velocity is $-U \ihat{A},$ and thus \eqref{eq:v_ca} implies that, for small angles,
\begin{align}
    \alpha
        \isdef
            \theta + \frac{\dot h}{U} -  \frac{d \dot \theta}{U}.
    \label{eq:alpha}
\end{align}
The lift coefficient is modeled as
\begin{align}
    C_\rmL(\alpha, \delta)
        &=
            C_{\rmL0} + C_{\rmL\alpha} \alpha
            +
            C_{\rmL\delta} \delta,
    \label{eq:CL_model}
\end{align}
where $C_{\rmL0}$ is the lift coefficient at $\alpha=0$ and $\delta=0,$ $C_{\rmL\alpha}$ is the lift-curve slope, and $C_{\rmL\delta}$ is the flap lift effectiveness.
The drag coefficient is modeled using a parasitic-plus-induced form
\begin{align}
    C_\rmD(\alpha,\delta)
        &=
            C_{\rmD0} + k_{\rmi} C_\rmL(\alpha, \delta)^2,
    \label{eq:CD_model}
\end{align}
where $C_{\rmD0} \ge 0$ is the zero-lift drag coefficient and $k_{\rmi} \ge 0$ is the induced-drag factor.
The pitching-moment coefficient is modeled as
\begin{align}
    C_\rmM(\alpha,\dot\theta, \delta)
        &=
            C_{\rmM 0} + C_{\rmM\alpha} \alpha
            + C_{\rmM \dot \theta} \frac{c}{2U}\dot\theta
            +
            C_{\rmM\delta} \delta,
    \label{eq:CM_model}
\end{align}
where $C_{\rmM0}$ is the moment coefficient at $\alpha=0,$ $\dot\theta=0,$ and $\delta=0,$ $C_{\rmM\alpha}$ is the moment-curve slope, $C_{\rmM \dot \theta}$ is the pitch-rate damping coefficient, and $C_{\rmM\delta}$ is the flap moment effectiveness.
% \todo{A moment about the aerodynamic center is independent of $\alpha$ by definition, but $C_{{\rm M}\alpha}=1.49$. Check the BACT reference point.}

\subsection{Dynamics}
Newton's second law $m \vect a_{c_\rmg/w/\rmA} = \vect f_{\rm total}$ yields
\begin{align}
    - m\ddot h \sin \theta
        &=
            -D - (m g - k_h\tilde h )\sin \theta + f_\rmr \cos \theta,
    \label{eq:hddotSin}
    \\
    m \ddot h\cos \theta
        &=
            - L + (m g - k_h\tilde h ) \cos \theta + f_\rmr \sin \theta.
    \label{eq:hddotCos}
\end{align}
Adding $-\sin \theta$ times \eqref{eq:hddotSin} and $\cos \theta$ times \eqref{eq:hddotCos} yields
\begin{align}
    m\ddot h
        &=
            D \sin \theta
            - L \cos \theta
            + m g - k_h\tilde h.
    \label{eq:hddot}
\end{align}
Next, Euler's equation yields
\begin{align}
    I_{yy/c_\rmg} \ddot \theta
        &=
            M_\rma -
            \kappa \tilde \theta
            +
            d L.
    \label{eq:Eulersddot}
\end{align}
Collecting the inertial and spring terms on the left-hand side and the aerodynamic and gravity terms on the right-hand side yields
\begin{align}
    m\ddot h + k_h\tilde h
        &=
            mg +
            D \sin \theta
            - L \cos \theta,
    \label{eq:hddot_2}
    \\
    I_{yy/c_\rmg} \ddot \theta + \kappa \tilde \theta
        &=
            M_\rma
            +
            d L.
    \label{eq:Eulersddot_2}
\end{align}

\subsection{State-Space Formulation}
Define the generalized coordinate vector
\begin{align}
    \eta
        \isdef
            \matl
                h \\
                \theta
            \matr.
\end{align}
The dynamics \eqref{eq:hddot_2}--\eqref{eq:Eulersddot_2} can then be written in second-order vector form as
\begin{align}
    M \ddot{\eta} + K \tilde{\eta} = F(\eta, \dot{\eta}),
    \label{eq:compact_eom}
\end{align}
where $\tilde{\eta} \isdef [ \tilde h \;\; \tilde\theta ]^\rmT$ denotes the displacement relative to the relaxed spring positions,
\begin{align}
    M
        &\isdef
            \matl
                m & 0 \\
                0 & I_{yy/c_\rmg}
            \matr,
    \qquad
    K
        \isdef
            \matl
                k_h & 0 \\
                0 & \kappa
            \matr,
    \label{eq:MK_def}
\end{align}
and the generalized force vector is
\begin{align}
    F(\eta, \dot{\eta})
        \isdef
            \matl
                mg + D(\eta,\dot{\eta}) \sin \theta - L(\eta,\dot{\eta}) \cos \theta \\
                M_\rma(\eta,\dot{\eta}) + d L(\eta,\dot{\eta})
            \matr.
    \label{eq:F_def}
\end{align}
The dependence of $L,$ $D,$ and $M_\rma$ on $\delta$ is omitted from the notation in \eqref{eq:F_def}.
Next, define the state vector
\begin{align}
    x
        \isdef
            \matl
                \eta \\ \dot \eta
            \matr
        =
            \matl
                h \\
                \theta \\
                \dot h \\
                \dot \theta
            \matr.
\end{align}
Then, the dynamics can be written as
\begin{align}
    \dot x
        =
            \matl
                \dot h \\
                \dot\theta \\
                \dfrac{1}{m}\!\left(mg + D\sin\theta - L\cos\theta - k_h\tilde h\right) \\
                \dfrac{1}{I_{yy/c_\rmg}}\!\left(M_\rma + dL - \kappa \tilde\theta\right)
            \matr,
    \label{eq:xdot}
\end{align}
where $L,$ $D,$ and $M_\rma$ are given by \eqref{eq:L_param}--\eqref{eq:M_param}.

\subsection{Parameters}
\label{sec:parameters}
Table~\ref{tab:parameters} lists the parameter values used in the stability analysis and the numerical simulations.
The values are adapted from the NASA Benchmark Active Controls Technology model \cite{waszak1998modeling}.

\begin{table}[ht!]
    \centering
    \caption{Model parameters.}
    \label{tab:parameters}
    \begin{adjustbox}{width=\textwidth}
    \begin{tabular}{c l c c | c l c c}
        \hline\hline
        Symbol & Description & Value & Units & Symbol & Description & Value & Units \\
        \hline
        $m$ & mass & 88.73 & kg &
        $g$ & acceleration due to gravity & 9.81 & m/s$^2$ \\
        $I_{yy/c_\rmg}$ & pitch inertia & 3.80 & kg$\cdot$m$^2$ &
        $C_{\rmL 0}$ & lift coefficient at $\alpha = 0$ & 0 & -- \\
        $k_h$ & plunge stiffness & 39199 & N/m &
        $C_{\rmL\alpha}$ & lift-curve slope & 4.584 & $\rm rad\inv$ \\
        $\kappa$ & pitch stiffness & 4067 & N$\cdot$m/rad &
        $C_{\rmL\delta}$ & flap lift effectiveness & 6.3$^\ast$ & $\rm rad\inv$ \\
        $h_0$ & relaxed length of linear spring & 0 & m &
        $C_{\rmD 0}$ & zero-lift drag coefficient & 0.0085 & -- \\
        $\theta_0$ & relaxed angle of torsional spring & 0 & rad &
        $k_\rmi$ & induced-drag factor & 0.035 & -- \\
        $d$ & distance from $c_\rmg$ to $c_\rma$ & 0.0508 & m &
        $C_{\rmM 0}$ & moment coefficient at $\alpha = 0$ & 0 & -- \\
        $S$ & planform area & 0.330 & m$^2$ &
        $C_{\rmM\alpha}$ & moment-curve slope & 1.490 & $\rm rad\inv$ \\
        $c$ & chord length & 0.405 & m &
        $C_{\rmM\dot\theta}$ & pitch-rate damping coefficient & $-2.6505$ & -- \\
        $\rho$ & air density & 1.225 & kg/m$^3$ &
        $C_{\rmM\delta}$ & flap moment effectiveness & $-0.246^\ast$ & $\rm rad\inv$ \\
        \hline\hline
    \end{tabular}
    \end{adjustbox}

    \vspace{0.5ex}
    {\footnotesize $^\ast$Ten times the value in \cite{waszak1998modeling}, increased to raise the control input effectiveness in this study.}
\end{table}
% \todo{Confirm the source of $C_{{\rm M}\dot\theta}$. The code takes it from C\_m\_alphadot.}
% \todo{Confirm $d$. The previous table listed $l=-0.175c$.}

\section{Linearization-based Stability Analysis}
\label{sec:stability_analysis}

To assess the local stability properties of the nonlinear aeroelastic system, we linearize the state-space model in \eqref{eq:xdot} about an equilibrium operating point corresponding to a fixed freestream velocity $U$.

Let $x^\ast$ denote an equilibrium point satisfying
$f(x^\ast) = 0$
where $f(x)$ is defined by \eqref{eq:xdot} with the aerodynamic forces given by \eqref{eq:L_param}--\eqref{eq:M_param} and the models in \eqref{eq:CL_model}--\eqref{eq:CM_model}. 
Linearizing the nonlinear dynamics about $x^\ast$ yields the first-order approximation $\Delta \dot x 
        = 
        A \, \Delta x,$ where 
$\Delta x \isdef x - x^\ast$ and 
\begin{align}
    A 
        \isdef 
        \left.\frac{\partial f}{\partial x}\right|_{x = x^\ast},
    \label{eq:linearized_system}
\end{align}

The local stability of the equilibrium is determined by the eigenvalues of the Jacobian matrix $A$. In particular, the system is locally asymptotically stable if all eigenvalues of $A$ lie in the open left-half complex plane, and instability (flutter) is identified when one or more eigenvalues cross into the right-half plane as the freestream velocity $U$ is varied.

\subsection{Linearization About the Nominal Equilibrium}

To make the dependence of the linearized dynamics on the freestream velocity explicit, consider the nominal equilibrium $x^\ast = 0,$
corresponding to
$    h = h_0 + mg/k_h, 
    \theta = 0, 
    \dot h = 0, 
$ and $
    \dot\theta = 0.$
At this equilibrium,
$\alpha = 0,
    \sin\theta = 0,
    $ and $
    \cos\theta = 1.$
Then, $A$ in \eqref{eq:linearized_system} is given by
\begin{align}
    A(U)
    =
    \begin{bmatrix}
        0 & 0 & 1 & 0 \\
        0 & 0 & 0 & 1 \\
        -\dfrac{k_h}{m} & a_{32}(U) & a_{33}(U) & a_{34}(U) \\
        0 & a_{42}(U) & a_{43}(U) & a_{44}(U)
    \end{bmatrix},
    \label{eq:Au}
\end{align}
where
\begin{gather}
    a_{32}(U)
    =
    \frac{\bar q}{m}
    \left(
        C_{\rmD0}
        -
        C_{\rmL\alpha}
    \right),
    \quad
    a_{33}(U)
    =
    -\frac{\bar q}{mU}C_{\rmL\alpha},
    \quad
    a_{34}(U)
    =
    \frac{\bar q d}{mU}C_{\rmL\alpha},
    \\
    a_{42}(U)
    =
    \frac{1}{I_{yy/c_\rmg}}
    \left(
        \bar q c C_{\rmM\alpha}
        +
        \bar q d C_{\rmL\alpha}
        -
        \kappa
    \right),
    \quad
    a_{43}(U)
    =
    \frac{\bar q}{I_{yy/c_\rmg}U}
    \left(
        cC_{\rmM\alpha}
        +
        dC_{\rmL\alpha}
    \right),
    \\
    a_{44}(U)
    =
    \frac{\bar q}{I_{yy/c_\rmg}U}
    \left(
        \frac{c^2}{2}C_{\rmM\dot\theta}
        -
        cdC_{\rmM\alpha}
        -
        d^2C_{\rmL\alpha}
    \right),
\end{gather}
and
\begin{align}
    \bar q \isdef \frac{1}{2}\rho U^2 S.
\end{align}

The structure of $A(U)$ reveals two distinct aerodynamic effects.
The terms proportional to $\bar q \sim U^2$ act as stiffness-like contributions, while terms proportional to $\bar q/U \sim U$ introduce velocity-dependent damping and coupling.
Note that the structural model in \eqref{eq:hddot_2}--\eqref{eq:Eulersddot_2} does not include explicit mechanical damping terms of the form $c_h \dot h$ or $c_\theta \dot\theta$.
Consequently, any damping in the linearized system arises solely from the aerodynamics.
In particular, the dependence of the aerodynamic forces on $\dot h$ and $\dot\theta$ introduces velocity-dependent terms that act as effective damping.
For sufficiently small nonzero freestream velocity, these aerodynamic contributions provide net positive damping, placing the eigenvalues of the linearized system in the open left-half plane.
As $U$ increases, the aerodynamic coupling reduces the effective damping of a conjugate eigenvalue pair.
\textit{Flutter} is the resulting oscillatory instability, whose onset corresponds to this pair crossing the imaginary axis into the open right-half plane.

\subsection{Parametric Sensitivity}
% To investigate robustness with respect to modeling uncertainty, selected physical and aerodynamic parameters appearing in \eqref{eq:MK_def} and \eqref{eq:F_def}, including mass $m$, stiffnesses $k$ and $\kappa$, air density $\rho$, and aerodynamic coefficients such as $C_{\rmM\alpha}$ and $C_{\rmL\alpha},$ are scaled over representative ranges. 
% For each parameter variation, the Jacobian in \eqref{eq:linearized_system} is recomputed, and its eigenvalues are evaluated.
To investigate the effect of aerodynamic uncertainty on the flutter boundary, we examine the effect of scaling the lift coefficient $C_{\rmL\alpha}$ in the linearized model. 
The coefficient $C_{\rmL\alpha}$ directly influences both the aerodynamic stiffness and the velocity-dependent aerodynamic damping terms in \eqref{eq:Au}, and therefore plays a central role in determining the onset of aeroelastic instability.

A scaling parameter $\gamma$ is introduced such that
$C_{\rmL\alpha}^{\rm scaled}
    =
    \gamma C_{\rmL\alpha},$
where $\gamma \in [0.5,\,1.5]$. 
For each value of $\gamma$, the Jacobian matrix $A(U)$ in \eqref{eq:Au} is recomputed over a range of freestream velocities, and the corresponding eigenvalues are evaluated.
The linearized system has four eigenvalues that appear as two complex-conjugate pairs. 
Let $\lambda_{1,2}
        =
            \sigma_1 \pm j\omega_1,$
            and
$\lambda_{3,4}
        =
            \sigma_2 \pm j\omega_2,$
where $\sigma_1,\sigma_2 \in \mathbb{R}$ denote the corresponding real parts. 
Since each conjugate pair shares the same real part, the stability analysis focuses on the evolution of $\sigma_1$ and $\sigma_2$ as functions of freestream velocity.

\begin{figure}[h]
    \begin{subfigure}{0.5\columnwidth}
        \centering
        \includegraphics[width=\linewidth]{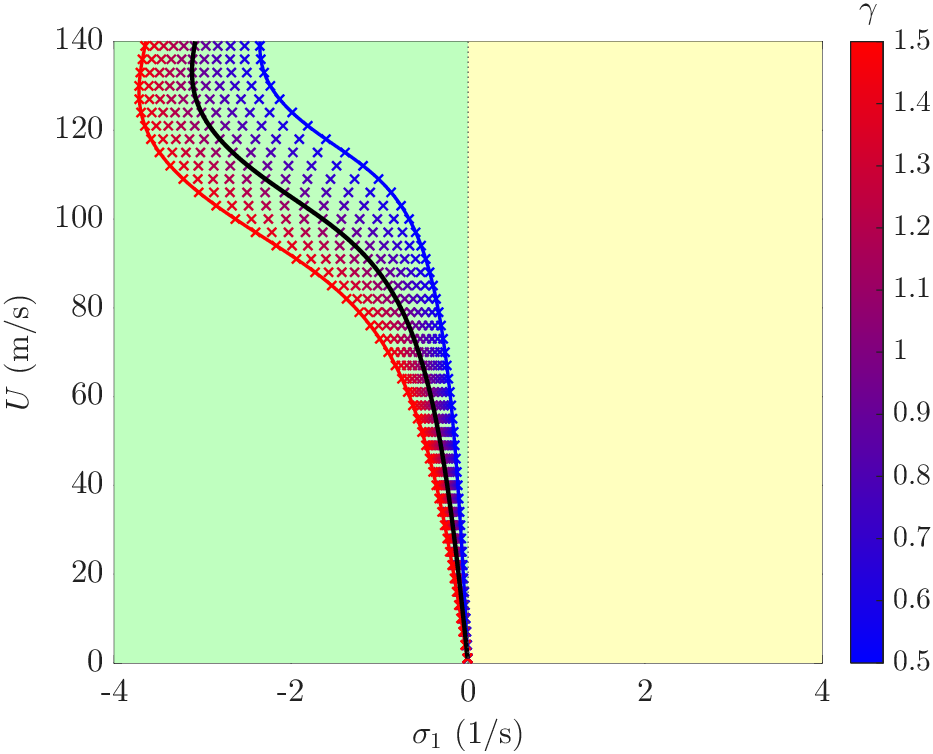}
        \caption{Variation of $\sigma_1$ with freestream velocity under scaling of $C_{{\rm L}\alpha}$.}
        \label{fig:Eig_sigma1}
    \end{subfigure}%
    \begin{subfigure}{0.5\columnwidth}
        \centering
        \includegraphics[width=\linewidth]{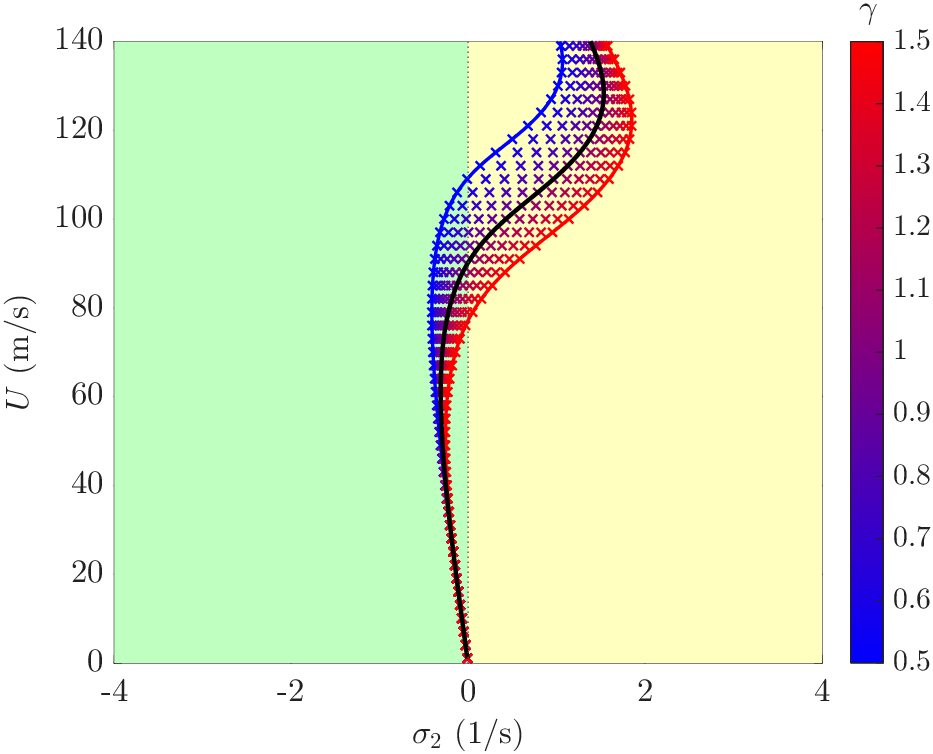}
        \caption{Variation of $\sigma_2$ with freestream velocity under scaling of $C_{{\rm L}\alpha}$.}
        \label{fig:Eig_sigma2}
    \end{subfigure}
    \caption{Evolution of the real parts of the eigenvalues of the linearized aeroelastic system as the freestream velocity $U$ and the $C_{{\rm L}\alpha}$ scaling parameter $\gamma$ are varied.}
    \label{fig:U0vsEig_Clalpha}
\end{figure}

Figure \ref{fig:U0vsEig_Clalpha} shows the evolution of the real parts of the eigenvalues as the freestream velocity increases. 
The color of each trajectory indicates the scaling parameter $\gamma$, ranging from blue ($\gamma = 0.5$) to red ($\gamma = 1.5$), with the black trajectory corresponding to the nominal case $\gamma = 1$. 
The green and yellow shaded regions denote the open left-half plane and open right-half plane, respectively.
These results illustrate the coupled influence of freestream velocity and aerodynamic lift sensitivity on the onset of flutter. 

For sufficiently small values of $C_{\rmL\alpha}$ and low freestream velocities, the real parts $\sigma_1$ and $\sigma_2$ remain negative, corresponding to locally stable dynamics. 
However, as either the freestream velocity or the lift coefficient $C_{\rmL \alpha}$ increases, the aerodynamic coupling modifies the effective damping and stiffness of the system, eventually causing one of the conjugate eigenvalue pairs to cross into the open right-half plane. 
Consequently, for any fixed freestream velocity, there exists a sufficiently large value of $C_{\rmL\alpha}$ that leads to flutter, and similarly, for any fixed value of $C_{\rmL\alpha}$, there exists a sufficiently large freestream velocity that destabilizes the coupled pitch--plunge dynamics. 
These results highlight the fundamental interplay between aerodynamic loading and structural dynamics in determining the aeroelastic stability boundary.

% The resulting eigenvalue trajectories provide insight into the sensitivity of the aeroelastic stability boundary to uncertain parameters and identify those quantities that most strongly influence the onset of instability. 
% Figure \ref{fig:U0vsEig_Clalpha} shows the variation of real part of the eigenvalues with respect to the freestream velocity. As previously mentioned, there is a scaling factor applied to the $C_{L_\alpha}.$ along the range of 0.5 to 1.5.
% This gain is represented by a color change from blue to red, with blue corresponding to the minimum scalor factor being applied. 
% The open left half plane and the open right half plane are denoted by the colored backgrounds of green and yellow, respectively. 
% The green background signifies negative $\sigma_1$ and $\sigma_2$ values which represent stability in the system. The yellow background demonstrates positve $\sigma_1$ and $\sigma_2$ values, indicating instability in the system. Furthermore, traces are present to indicate all $\sigma_1$ and $\sigma_2$ values which are scaled by the same constant, $\alpha$. In these figures, the blue line represents all the eigenvalues for $\alpha$ equals 0.5, followed by black for which $\alpha$ = 1 and concluding with pink where $\alpha$ = 1.5.

% 

% \begin{figure}[H]

%     \centering
%     \includegraphics[width=3in]{BACT_Figures/U0vsEig_m.eps}
%     \caption{$m$}
%     \label{fig:placeholder}

% \end{figure}

% \begin{figure}[H]

%     \centering
%     \includegraphics[width=3in]{BACT_Figures/U0vsEig_Ktheta.eps}
%     \caption{$K_\theta$}
%     \label{fig:placeholder}

% \end{figure}

% \begin{figure}[H]

%     \centering
%     \includegraphics[width=3in]{BACT_Figures/U0vsEig_Kh.eps}
%     \caption{$K_h$}
%     \label{fig:placeholder}

% \end{figure}

% \begin{figure}[H]

%     \centering
%     \includegraphics[width=3in]{BACT_Figures/U0vsEig_rho.eps}
%     \caption{$\rho$}
%     \label{fig:placeholder}

% \end{figure}

% \begin{figure}[H]

%     \centering
%     \includegraphics[width=3in]{BACT_Figures/U0vsEig_Cmalphadot.eps}
%     \caption{$C_{M_{\dot{\alpha}}}$}
%     \label{fig:placeholder}

% \end{figure}

% \begin{figure}[H]

%     \centering
%     \includegraphics[width=3in]{BACT_Figures/U0vsEig_Clalpha.eps}
%     \caption{$C_{l_\alpha}$}
%     \label{fig:placeholder}

% \end{figure}

\section{Data-Driven Control}
\label{sec:control}
The linearization-based analysis in Section~\ref{sec:stability_analysis} highlights the sensitivity of the aeroelastic system to operating conditions and parametric uncertainty, motivating the need for adaptive control strategies that do not rely on an accurate model.
In this section, we investigate the use of Dynamic Mode Adaptive Control (DMAC) to suppress vibration in the pitch--plunge testbed described by \eqref{eq:xdot}.
DMAC is a data-driven control framework that constructs a local linear representation of the system dynamics online and synthesizes feedback control directly from measured state trajectories.
A detailed description of DMAC, including the recursive identification algorithm and LQR controller synthesis, can be found in \cite{oveissi2026dmac}.

At each sampling instant $k$, DMAC recursively identifies a local linear model of the measured dynamics of the form
\begin{align}
    \xi_{k+1}
    =
    \Theta_k^{\rmS}
    \begin{bmatrix}
        \xi_k \\
        u_k
    \end{bmatrix},
\end{align}
where $\xi_k$ is the measured portion of the state and $\Theta_k^{\rmS}$ denotes the identified state-space model obtained using the Dynamic Mode Approximation (DMA) algorithm.
The identified model is updated recursively using streaming measurements, thereby allowing the controller to adapt to changing operating conditions and unmodeled dynamics.

The identified model is subsequently used to synthesize the feedback controller online.
In particular, the controller gain matrix $\Theta_k^{\rmC}$ is computed using a discrete-time linear quadratic regulator (LQR) design augmented with integral action to enable command following and flutter suppression.
The control signal is given by
\begin{align}
    u_k
        =
            \Theta_k^\rmC
            \matl
                \xi_k \\
                q_k
            \matr
            +
            v_k,
\end{align}
where the integrator state $q_k$ satisfies $q_{k+1} = q_k + T_\rms (r - y_k),$ $T_\rms$ is the sampling interval, $r$ is the commanded plunge, $y_k$ is the measured plunge, and $v_k \sim \SN(0,\sigma_v^2)$ is a zero-mean Gaussian excitation signal.
The excitation signal provides additional excitation for online identification.
The resulting controller combines online data-driven system identification with LQR feedback synthesis.

\section{Numerical Simulations}
\label{sec:numerical}
In this section, we apply DMAC to the pitch--plunge testbed described by \eqref{eq:xdot}.
We first describe the simulation setup, then show the open-loop response and command following above the flutter speed, and finally consider a varying freestream velocity.

\subsection{Simulation Setup}
\label{sec:simulation_setup}
The nonlinear model \eqref{eq:xdot} is simulated with the parameters listed in Table~\ref{tab:parameters}.
The control input is the trailing-edge flap deflection $\delta,$ which is limited to $\pm 30$~deg.
The objective is to suppress flutter while following a commanded plunge displacement.
The initial condition is $h(0) = 0.03$~m, $\theta(0) = 0.1$~rad, $\dot h(0) = 0,$ and $\dot \theta(0) = 0.$

In DMAC, the measured state is $\xi_k = x_k,$ the output is $y_k = h_k,$ and the sampling interval is $T_\rms = 0.01$~s.
The forgetting factor is $\lambda = 0.9995,$ the initial identified model is zero, and the initial covariance is $0.01 I.$
The LQR weights used in DMAC are $Q = 10 I_5$ and $R = 10.$

To limit parameter drift, the identification update is terminated based on the performance of the closed-loop system, as described in \cite{salazar2026performancebased}.
Specifically, we compute an exponentially weighted root-mean-square value $z_{\rm rms}$ of the tracking error with forgetting factor $0.99.$
The identified model and the covariance are updated when $z_{\rm rms}$ is at least $3$~mm, and are held constant otherwise.
%
% Updates resume if $z_{\rm rms}$ rises above this threshold.

The standard deviation of the excitation signal is also scheduled using $z_{\rm rms}$ as
% \begin{align}
    $\sigma_{v,k}
        =
            \bar \sigma_v \sqrt{1 - e^{-z_{{\rm rms},k}/\bar z}},$
%     \label{eq:sigma_v}
% \end{align}
where $\bar \sigma_v = 0.1$~rad and $\bar z = 0.15$~m.
The excitation thus decreases as the tracking error decreases.

DMAC is compared with an LQR whose constant gain is designed using the linearization of \eqref{eq:xdot} about the equilibrium $x^\ast$ at $U = 122$~m/s.
The gain minimizes a continuous-time quadratic cost under sampled-data control with sampling interval $T_\rms.$
The state is augmented with the integrator state, and the state and input weights are ${\rm diag}(10,10,1,1,100)$ and $10,$ respectively.
LQR uses the same integrator and flap limit as DMAC.

\subsection{Open-Loop Response}
\label{sec:open_loop}
\begin{figure}[ht!]
    \centering
    \includegraphics[width=\linewidth]{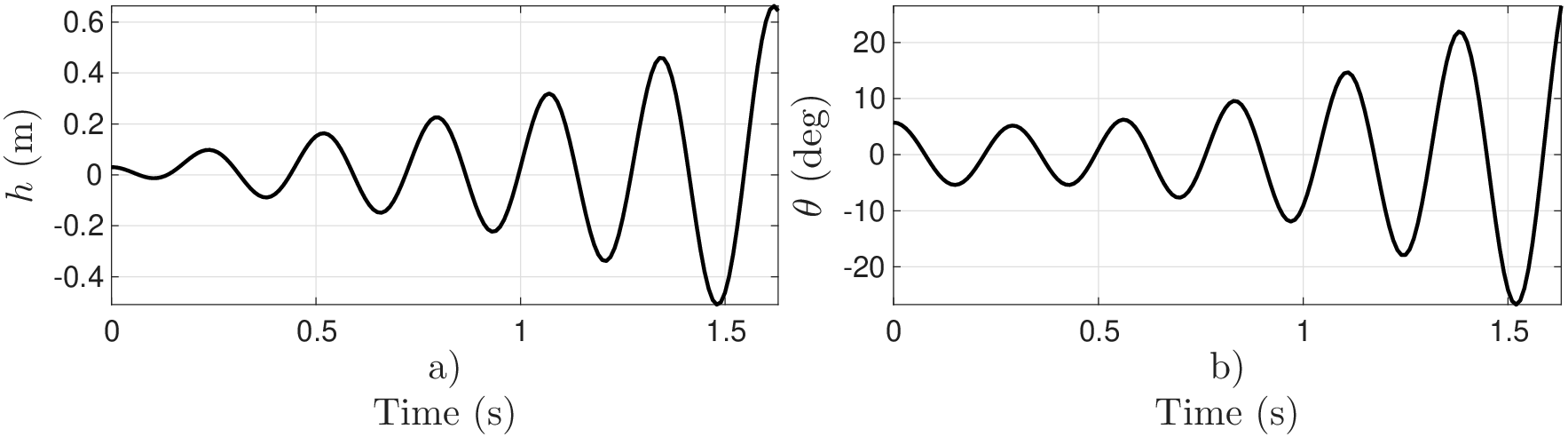}
    \caption{Open-loop response at $U = 122$~m/s. (a) Plunge displacement and (b) pitch angle.}
    \label{fig:OL_response}
\end{figure}

The freestream velocity is set to $U = 122\ {\rm m/s}$.
At this velocity, the open-loop pitch--plunge system is unstable, as indicated by the stability analysis in Section~\ref{sec:stability_analysis}.
Figure~\ref{fig:OL_response} shows the open-loop response of the pitch--plunge testbed, where (a) shows the plunge displacement and (b) shows the pitch angle.
The open-loop system exhibits growing oscillatory behavior in both plunge and pitch.

\subsection{Command Following Above the Flutter Speed}
\label{sec:command_following}
\begin{figure}[ht!]
    \centering
    \includegraphics[width=\linewidth]{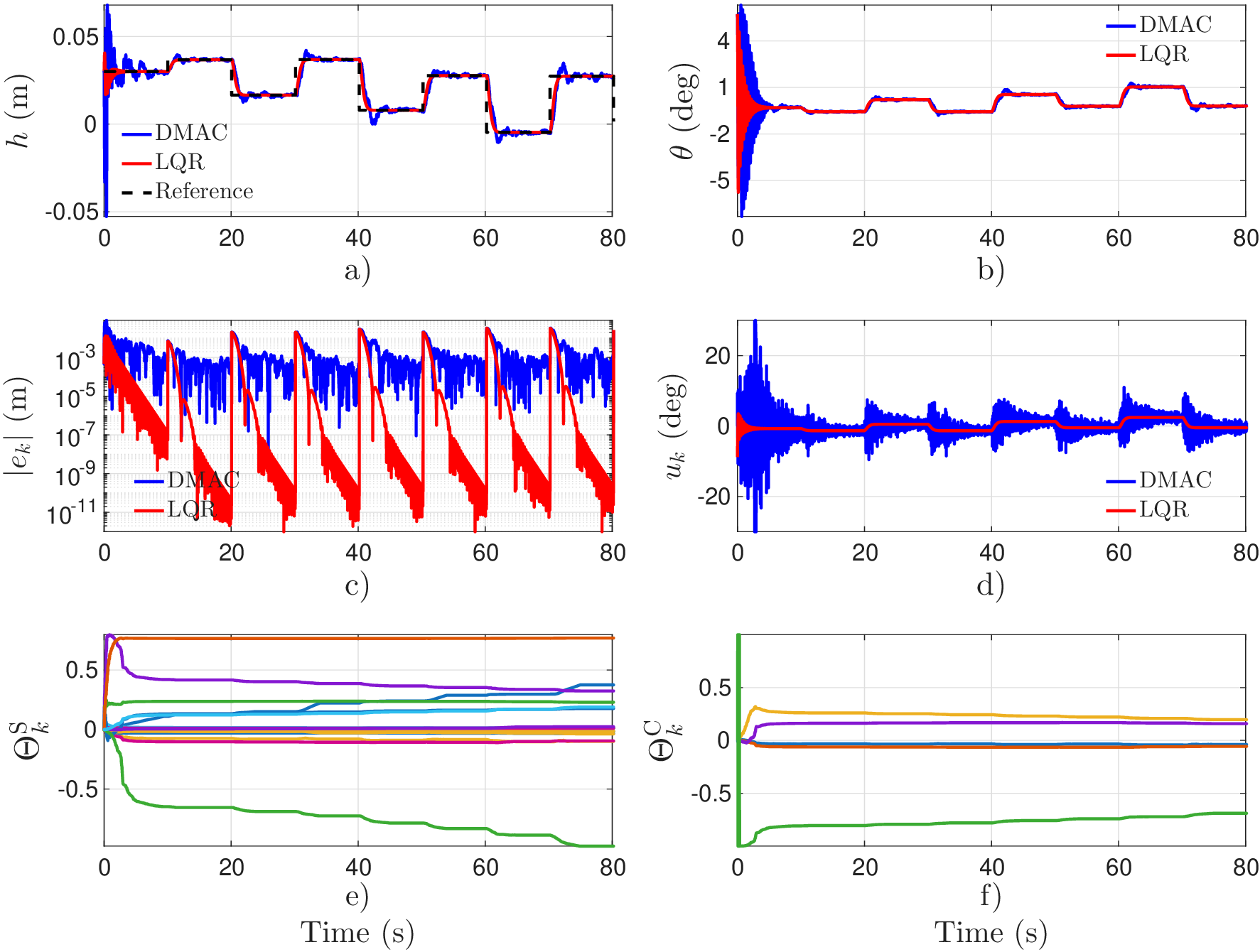}
    \caption{Command following at $U = 122$~m/s. (a) Plunge displacement, (b) pitch angle, (c) absolute tracking error, (d) flap deflection, (e) identified model, and (f) controller gain.}
    \label{fig:DMAC_122}
\end{figure}

The freestream velocity is set to $U = 122\ {\rm m/s}$, and the commanded plunge displacement is initially $0.03\ {\rm m}$ and then changes every $10\ {\rm s}$ to a value drawn uniformly from $[-0.1c, 0.1c]$.
Figure~\ref{fig:DMAC_122} shows the response of the pitch--plunge testbed with DMAC and LQR for command following above the flutter speed.
In particular,
(a) shows the plunge displacement,
(b) shows the pitch angle,
(c) shows the absolute tracking error on a logarithmic scale,
(d) shows the control input, and
(e) and (f) show the evolution of the identified model $\Theta_k^\rmS$ and controller gains $\Theta_k^\rmC$ in DMAC, respectively.

The DMAC controller suppresses the oscillations and tracks the commanded plunge displacement.
During the initial transient, the response is oscillatory and the control input reaches its limit.
Following each reference change, the controller adjusts its control action and drives the plunge displacement toward the new reference.
As expected, LQR achieves better performance than DMAC, since the LQR gains are computed with knowledge of the model and all of the aerodynamic and control derivatives.
In contrast, DMAC uses only the streaming measurements, and, after the initial transient, its plunge response is close to that of LQR, although its tracking error remains larger.
The remaining states stay bounded throughout the maneuver.

\subsection{Varying Freestream Velocity}
\label{sec:speed_switching}
\begin{figure}[ht!]
    \centering
    \includegraphics[width=\linewidth]{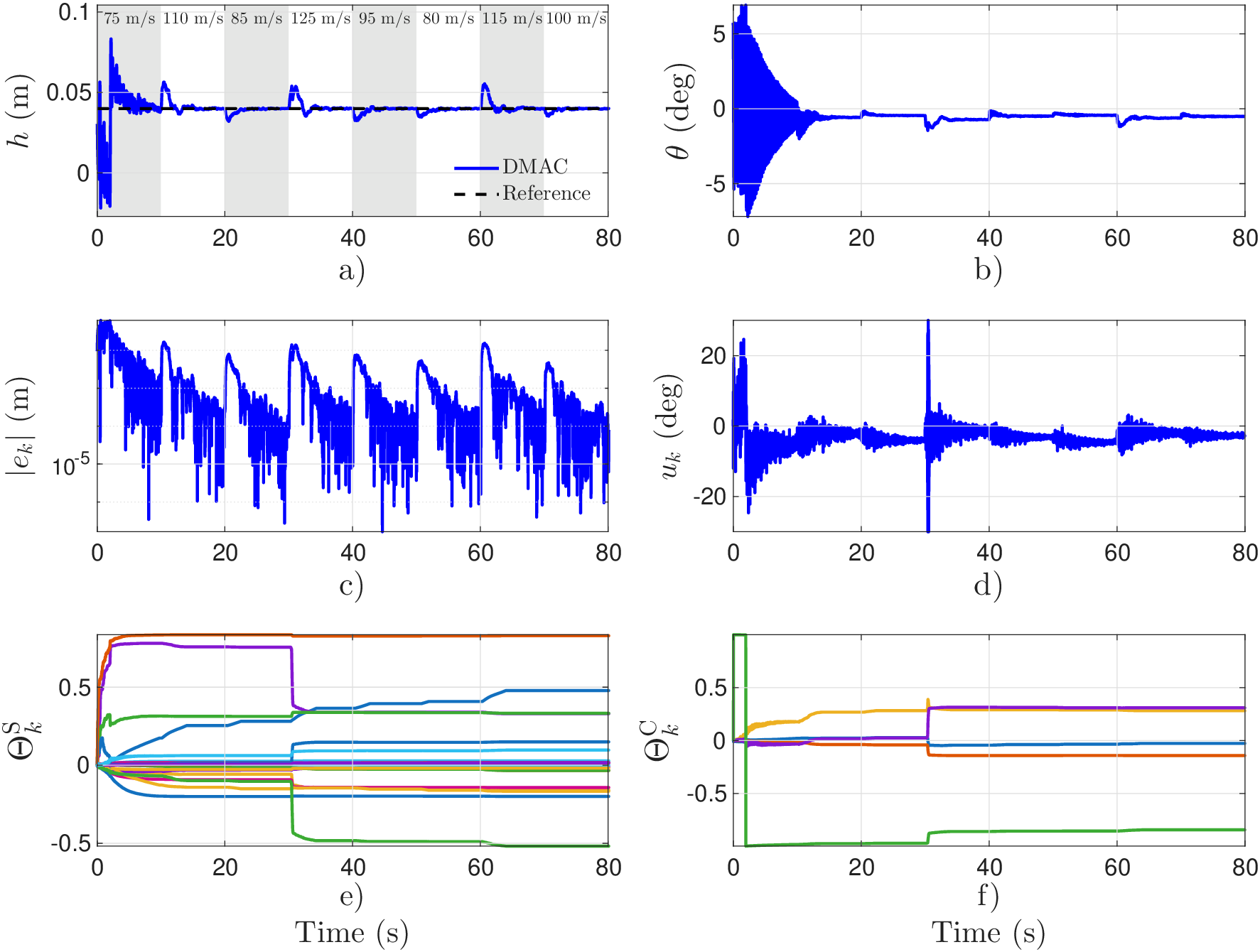}
    \caption{Closed-loop response of the pitch--plunge testbed with DMAC for flutter suppression and reference tracking under varying freestream velocity.}
    \label{fig:DMAC_switching}
\end{figure}

The commanded plunge displacement is set to $0.04\ {\rm m}$, and the freestream velocity is changed every $10\ {\rm s}$ between $75\ {\rm m/s}$ and $125\ {\rm m/s}$, crossing the flutter speed several times.
The DMAC controller is neither reset nor retuned when the velocity changes.
Figure~\ref{fig:DMAC_switching} shows the response of the pitch--plunge testbed with DMAC for flutter suppression and reference tracking under varying freestream velocity.
In particular,
(a) shows the plunge displacement, where the shaded bands indicate the freestream velocity,
(b) shows the pitch angle,
(c) shows the absolute tracking error on a logarithmic scale,
(d) shows the control input, and
(e) and (f) show the evolution of the identified model $\Theta_k^\rmS$ and controller gains $\Theta_k^\rmC$, respectively.

The DMAC controller suppresses the oscillations and tracks the commanded plunge displacement across the range of freestream velocities.
During the initial transient, the response is oscillatory.
Following each change in velocity, the controller adjusts its control action and drives the plunge displacement back to the reference.
After the initial transient, the largest transients follow the changes to $110$, $125$, and $115\ {\rm m/s}$.
Following the change to $125\ {\rm m/s}$, the control input reaches its limit, and the identified model and controller gains change abruptly.
Note that DMAC is not provided with the freestream velocity or the model, and adapts to each change using only the streaming measurements.
The remaining states stay bounded throughout the maneuver.

\section{Conclusion}
\label{sec:conclusions}
This paper presented a first-principles nonlinear model of a pitch--plunge aeroelastic system that captures the coupled interaction between structural dynamics and aerodynamic forces.
A linearization-based stability analysis was used to characterize the onset of flutter and investigate the sensitivity of the stability boundary to freestream velocity and the lift-curve slope.
The results demonstrated the influence of aerodynamic coupling and operating conditions on the stability of the pitch--plunge system.

Dynamic Mode Adaptive Control (DMAC) was employed for flutter suppression and plunge tracking using online local system identification and LQR feedback synthesis without requiring a model of the system dynamics.
Numerical simulations demonstrated that DMAC stabilizes the aeroelastic system beyond the flutter boundary and tracks plunge commands.
As expected, LQR, which is computed with knowledge of the model and the control derivatives, achieved a smaller tracking error, whereas DMAC achieved a plunge response close to that of LQR after the initial transient using only streaming measurements.
With the freestream velocity changing across the flutter speed, DMAC maintained command following without being reset or retuned.
These results demonstrate the potential of DMAC for data-driven flutter suppression and reference tracking in aeroelastic systems.

% 
% \printbibliography
\bibliography{b,Flutter,ref_aero}
\end{document}